\documentstyle[11pt]{article}
\newcommand{\BEQ}{\begin{equation}}  \newcommand{\WA}{{\cal WA}_{n-1}}
\newcommand{\BEA}{\begin{eqnarray}}  \newcommand{\EEA}{\end{eqnarray}}
\newcommand{\EEQ}{\end{equation}} 
                                  
\newcommand{\BAR}{\begin{array}}     \newcommand{\wt}{\sqrt{2}}
\newcommand{\EAR}{\end{array}}       
\newcommand{\vph}{\varphi}           \newcommand{\ra}{\rightarrow}
\newcommand{\sig}{\sigma}   \newcommand{\NIF}{N \rightarrow \infty}
\newcommand{\Ga}{\Gamma}    \newcommand{\la}{\lambda}
\newcommand{\NC}{\vph=\phi=0}      \newcommand{\SD}{\vph=\phi=\pi/2}
\newcommand{\pih}{\frac{\pi}{2}}   
\newcommand{\SK}{\sum_{k=1}^{n-1}} \newcommand{\pn}{\frac{\pi}{2n}}
\newcommand{\zeile}[1]{\vskip #1 \baselineskip} 
\newcommand{\hzeile}{\vskip 6 true pt}	
\font\EXTRA=cmss12 scaled \magstep2
\font\Extra=cmss12 scaled \magstep1
\font\extra=cmss10 scaled \magstep0
\font\extras=cmss10 scaled 750
\setbox2=\hbox{{{\EXTRA Z}}}
\setbox3=\hbox{{{\Extra Z}}}
\setbox4=\hbox{{{\extra Z}}}
\setbox5=\hbox{{{\extras Z}}}
\setbox6=\hbox{{{\extras z}}}
\def\BBZ{{{\EXTRA Z}}\hskip-\wd2\hskip 3.5 true pt{{\EXTRA Z}}}
\def\BZ{{{\Extra Z}}\hskip-\wd3\hskip 3 true pt{{\Extra Z}}}
\def\Z{{{\extra Z}}\hskip-\wd4\hskip 2.5 true pt{{\extra Z}}}

\def\BZED{\hbox{{\EXTRA\BBZ}}}
\def\ZED{\hbox{{\Extra\BZ}}}
\def\Zed{\hbox{{\extra\Z}}}

\begin{document}
\begin{titlepage}
\null   \begin{center}   \vskip 1cm
{\LARGE {\bf Multi-particle structure in the $\BZED_n$-chiral Potts models}}
\vskip 1.5cm     {\Large G.\ von Gehlen $\!{}^{*}$~and~
                         A.\ Honecker $\!{}^{**}$ }    \zeile{1}
{\large Physikalisches Institut der Universit\"{a}t Bonn \\
Nu{\ss}allee 12, D 5300 Bonn 1, Germany} \\    \zeile{3}   \end{center}
{\large \bf Abstract :}
We calculate the lowest translationally invariant levels of the
$\Zed_3$- and $\Zed_4$-symmetrical chiral Potts quantum chains,
using numerical diagonalization of the hamiltonian for
$N \leq 12$ and $N \leq 10$ sites, respectively, and extrapolating $\NIF$.
In the high-temperature massive phase we find that the pattern of the low-lying
zero momentum levels can be explained assuming the existence of $n-1$ particles
carrying $\Zed_n$-charges $Q\!=\!1, \ \dots, \ \! n-1$ (mass $m_Q$), and their
scattering states. In the superintegrable case the masses of the $n-1$
particles become proportional to their respective charges: $m_Q = Q m_1$.
Exponential convergence in $N$ is observed for the single particle gaps,
while power convergence is seen for the scattering levels.
We also verify that qualitatively the same pattern
appears for the self-dual and integrable cases.
For general $\Zed_n$ we show that the energy-momentum relations of the
particles show a parity non-conservation asymmetry which for
very high temperatures is exclusive due to the presence of
a macroscopic momentum $P_m=(1-2Q/n)/\phi$, where $\phi$ is the chiral angle
and $Q$ is the $\Zed_n$-charge of the respective particle. \vfill
\noindent
\noindent   BONN-HE-92-32        \hspace{88mm}  October 1992 \\
\noindent   hep-th/9210125 \\
${}^{*}$ \hspace*{1mm}   e-mail-address: unp02f@ibm.rhrz.uni-bonn.de \\
${}^{**}$  e-mail-address: unp06b@ibm.rhrz.uni-bonn.de
\end{titlepage}
\section{Introduction}     \par The $\Zed_3$-chiral
Potts model has been introduced in 1981 by Ostlund and Huse in order to
describe incommensurate phases of physisorbed systems \cite{Os81,Hu81}, e.g.
monolayer krypton on a graphite surface \cite{Ein}.
The phase structure of several versions of the model has been studied
by various methods: Mean field, Monte-Carlo, renormalization group,
transfer-matrix partial diagonalization and finite-size scaling of quantum
chains \cite{denis,sel,geh2,evr}. In the following we shall focus mostly
on the quantum chain version of the model. \par  In 1981-82 quantum chain
hamiltonians for the chiral Potts model were obtained \cite{MRR,cent}
via the $\tau$-continuum limit \cite{FrS,Kog}
from the Ostlund-Huse two-dimensional model.
These were not self-dual, so that the location of the critical manifold
was possible only by finite-size scaling \cite{geh2,ves}. In 1983
Howes, Kadanoff and DenNijs \cite{how} introduced a self-dual
$\Zed_3$-symmetrical chiral quantum chain, which, however, does not correspond
to a two-dimensional model with positive Boltzmann weigths. Therefore
in the self-dual model the immediate
connection to realistic physisorbed systems is lost, but its
peculiar mathematical structure has attracted much interest: Howes {\em et al.}
found that the lowest gap of the self-dual model is linear in the inverse
temperature, and subsequently in ref.\ \cite{geh} it was shown that the model
fulfils the Dolan-Grady integrability conditions \cite{dol}.
(now usually called "superintegrability" \cite{coy}). A whole series of
$\Zed_n$-symmetrical quantum chains has been defined, which satisfy the
superintegrability conditions \cite{geh}. The Dolan-Grady integrability
conditions have been shown to be equivalent to the Onsager algebra
\cite{ons,dav,rr}, which entails that all eigenvalues of the hamiltonian
have a simple Ising-type form. The well-known Ising quantum chain in a
transverse field is the $\Zed_2$-version of the superintegrable chiral Potts
models.
\par Much work has been done in the past few years in order to obtain
analytic expressions for the complete spectrum of the
$\Zed_3$-superintegrable model \cite{coy,ba1}. In the next Section we shall
quote some of these results. The aim of the present paper is not
to add to the analytic
calculations of the special superintegrable case, but rather to use numerical
finite-size analysis in order to investigate, how the integrable model is
embedded in the more general
versions of the chiral $\Zed_3$- and $\Zed_4$-models. We shall study the
low-lying levels of the excitation spectrum and investigate the possibility of
a particle interpretation.  \par
In the neighbourhood of conformal points of isotropic theories, very
simple particle patterns have been found by Zamolodchikov's perturbation
expansion \cite{zamo}, which is applicable to the non-chiral limit of
the chiral Potts-models. We shall follow these particle patterns by
varying the chiral angles, and try to find out which
special properties of the spectrum lead, for particular
parameter values, to superintegrability. Since for this purpose we have to
study the spectrum through a large probably non-integrable
parameter range, at present there is no alternative to numerical methods.
It turns out that even in those cases which can be solved analytically,
some basic features can be discovered rather easily through a numerical
calculation, because the exact formulae are quite involved.
\par The plan of this paper is as follows: In Sec.2 we start with the basic
definitions of the chiral Potts model for general $\Zed_n$-symmetry, which
then are specialized to the $\Zed_3$- and $\Zed_4$-cases.
Sec.3 collects some basic analytic results which are available for the generic
$\Zed_n$-symmetric model. In Sec.4 we give our detailed finite-size
numerical results which confirm the two-particle interpretation of the
low-lying spectrum of the self-dual version of the $\Zed_3$-model.
Sec.5 discusses the analogous results for the self-dual version of the
$\Zed_4$-model and shows that the low-lying spectrum is well described in terms
of three elementary particles. While up to this point, we consider
translationally invariant states only, in Sec.6 we present observations on the
energy-momentum dispersion relations of the elementary particles and the
effects of parity-violation on the spectrum. Finally, Sec.7 collects our
conclusions.

\section{The chiral $\ZED_n$-Potts quantum chain}
\par The chiral $\Zed_n$-symmetric Potts spin quantum chain \cite{geh} with $N$
sites is defined by the hamiltonian
\BEQ
H^{(n)} = -\sum_{j=1}^{N} \sum_{k=1}^{n-1} \left\{ \bar{\alpha}_k \, \sig_{j}^k
    + \la \, \alpha_k \, \Ga_{j}^k \, \Ga_{j+1}^{n-k} \right\}.    \label{H}
\EEQ
Here $\sig_j$ and $\Ga_j$ are $n\times n$-matrices acting at site $j$,
which satisfy the relations
\BEQ \sig_j\,\Ga_{j'} =\Ga_{j'}\,\sig_j~ \omega^{\delta_{j,j'}}, \hspace{18mm}
\sig_j^n=\Ga_j^n={\bf 1},\hspace{18mm}\omega =\exp(2\pi i/n).   \label{om} \EEQ
A convenient representation for $\sig$ and $\Ga$ in terms of diagonal and
lowering matrices, respectively, is given by
\BEQ  (\,\sig\,)_{l,m} = \delta_{l,m} \omega^{l-1},~~~~~~~~~~~~
 (\,\Ga\,)_{l,m} = \delta_{l+1,m}~~({\textstyle mod}~n). \label{sgr} \EEQ
We shall assume periodic boundary conditions: $\Ga_{N+1} = \Ga_1$.
\par
The model contains $2n-1$ parameters: the real inverse temperature $\la$
and the complex constants $\alpha_k$ and $\bar{\alpha}_k$. We shall only
consider the case of $H^{(n)}$ being hermitian:
$\alpha_k = \alpha_{n-k}^{\star}$
and $\bar{\alpha}_k = \bar{\alpha}_{n-k}^{\star}$.
For $n > 2$ and generic complex $\alpha_k$ and $\bar{\alpha}_k$, the
Perron-Frobenius theorem does not apply and the spectrum of $H^{(n)}$
may show ground-state level crossings.
$H^{(n)}$ commutes with the $\Zed_n$ charge operator
$\tilde{Q} = \prod_{j=1}^N \sig_j$.   The eigenvalues of $\tilde{Q}$ have the
form $\exp(2\pi i Q/n)$ with $Q$ integer. We shall refer to the $n$ charge
sectors of the spectrum of $H^{(n)}$ by $Q=0,\ldots,n-1$, respectively.
Parity is not a
good quantum number, but $H^{(n)}$ is translational invariant, so that
each eigenstate of $H^{(n)}$ can be labelled by its momentum eigenvalue $p$.
\par For $\alpha_k = \bar{\alpha}_k$ the model is self-dual
with respect to the reflection $\la \ra \la^{-1}$. If we choose \cite{geh}
\BEQ  \alpha_k = \bar{\alpha}_k = 1 - i \cot(\pi k/n)    \label{aint} \EEQ
then the model is "superintegrable" \cite{al2}, i.e. it fulfils the Dolan-Grady
\cite{dol}
conditions and therefore the $\la$-dependence of all eigenvalues $E(\la)$
of $H^{(n)}$ has the special Ising-like form \cite{dav,al2}:
\BEQ  E(\la) = a + b\la + \sum_j 4m_j\sqrt{1 +
 2\la\cos\theta_j +\la^2}.    \label{Isi}  \EEQ  Here $a, b$ and $\theta_j$
are real numbers. The $m_j$ take the values $m_j =-s_j,-s_j+1,\ldots,s_j$,
where $s_j$ is a finite integer. For details concerning formula (\ref{Isi}),
see \cite{dav,al2}.
\par Albertini {\em et al.} \cite{al2,au2} have obtained a spin chain of the
form (\ref{H}) with
\BEQ    \alpha_k = e^{i(2k/n-1)\phi}/\sin(\pi k/n),~~~~~~~
  \bar{\alpha}_k = e^{i(2k/n-1)\vph}/\sin(\pi k/n)           \label{alpha} \EEQ
\BEQ  \cos \vph = \la \cos \phi                         \label{int}   \EEQ
as a limiting case of an integrable two-dimensional lattice model. The
Boltzmann weights of their two-dimensional model do not have the usual
difference property \cite{au2} and satisfy a new type of
Yang-Baxter equations which involve spectral parameters defined on
Riemann surfaces of higher genus. So a quantum chain (\ref{H}) with
coefficients (\ref{alpha}), (\ref{int}) is integrable. The superintegrable
case is contained in (\ref{alpha}) for $\SD$.   \par
A number of analytic results is available for the superintegrable case.
Some of these will be reviewed in the next Section. Starting 1987, the
Stony-Brook-group has published a series of papers \cite{coy,al2,au2} which
give analytic calculations of the spectrum of the
$\Zed_3$-superintegrable quantum chain.
Recently, the completeness of the analytic
expressions for the levels of the superintegrable $\Zed_3$-model
has been shown by Dasmahapatra {\em et al.} \cite{DKCoy}.
\par   In the non-chiral limiting case $\NC$ of the hamiltonian (\ref{H}) with
coefficients (\ref{alpha}) we obtain the lattice version of the parafermionic
$\Zed_n$-symmetrical Fateev-Zamolodchikov-models $\WA$ \cite{fateev}.
These quantum chains are self-dual. At the self-dual point $\la=1$ and for
$\NIF$ they show an extended conformal symmetry
with central charge $c=2(n-1)/(n+2)$.
Exact solutions of these $\WA$-hamiltonian chains \cite{Al86,Alc,AlLi,FaLy}
have been obtained
through Bethe-ansatz techniques \cite{AlLi,Tsv88}. Very recently,
Cardy \cite{JC} has shown that a particular integrable perturbation of the
critical $\WA$-models leads to self-dual chiral Potts models.
\par  Since later we shall mostly study the $\Zed_3$- and $\Zed_4$-versions of
(\ref{H}), we now list how $H^{(n)}$ specializes for these cases. If we
keep insisting on hermiticity, $H^{(3)}$ depends on three parameters apart
from a normalization \cite{how}. In accordance with (\ref{alpha}) we write:
\BEQ    H^{(3)} = -\frac{2}{\sqrt{3}} \sum_{j=1}^{N} \left\{
 e^{-i\vph/3} \sig_{j} + e^{+i\vph/3} \sig^{+}_{j} + \la \left(
 e^{-i\phi/3} \Ga_{j}\Ga^{+}_{j+1} + e^{+i\phi/3}
 \Ga^{+}_{j}\Ga_{j+1} \right) \right\}.               \label{H3}      \EEQ
Here $\sig_{j}$ and $\Ga_{j}$ are $3\times 3$-matrices acting at site $j$:
\BEQ   \sig_j = \left( \BAR{ccc} 1&0&0\\0&\omega&0\\0&0&\omega^2
       \EAR \right)_{j},  \hspace{1cm} \Ga_j =
 \left( \BAR{ccc} 0&0&1\\1&0&0\\0&1&0 \EAR \right)_{j}  \label{sG}    \EEQ
and  $\omega = \exp(2\pi i/3)$.
We shall consider $0 \leq \la \leq \infty$;
$0 \leq \phi,\vph \leq \pi$ (for reflection properties of $H^{(3)}$ see
\cite{how}).        In order not to get
lost in three-dimensional diagrams, often we shall concentrate on
two cases in which there is one relation between the three parameters:
the integrable case (INT), where the three parameters are related by
eq.(\ref{int}), and the self-dual case (SD) where ~$\phi = \vph$. For generic
$\vph$ and $\la$,
the self-dual case is not known to be integrable. The SD case contains
the non-chiral limit $\NC$, in which we obtain the ${\cal WA}_2$-model, which
coincides with the $\Zed_3$-standard Potts model.
It has a second-order phase transition at $\la=1$, which for
$\NIF$ is described by a conformal field theory with central charge
$c=4/5$ \cite{Dot,FQS,GRR}.
\par The phase diagram of the SD-chiral $\Zed_3$-model shows four different
phases \cite{al2,GK,gehlenz3}, see Fig.~1: for small chiral angle $\vph$ we
have
oscillating massive high- and low-temperature phases at $\la<1$ and $\la>1$,
respectively, except for a small incommensurate phase interval around $\la =1$.
The two incommensurate phases appear centered around $\la=1$ and get wider
in $\la$ as $\vph$ increases.      \par The
hermitian $\Zed_4$-symmetric $H^{(4)}$ contains five parameters (again apart
from a normalization), which we denote by $\la$, $\phi$, $\vph$, $\beta$ and
$\tilde{\beta}$:    \BEQ
H^{(4)} = -\wt \sum_{j=1}^N \left\{
    e^{-i\vph/2}\sig_j +\beta\sig_j^2 + e^{i\vph/2}\sig_j^3 \right.
    + \la \left. \left\lbrack e^{-i\phi/2}
\Ga_j\Ga_{j+1}^3 +\tilde{\beta}\Ga_j^2\Ga_{j+1}^2 +e^{i\phi/2}\Ga_j^3 \Ga_{j+1}
    \right\rbrack \right\}        \nonumber       \\     \label{H4}
 \vspace*{3mm}\EEQ
where $\sig_j$ and $\Ga_j$ are now $4\times 4$-matrices
obeying (\ref{sgr}). For simplicity, and in agreement with (\ref{alpha}),
we shall consider only the case $\beta = \tilde{\beta}=1/\wt$, so that we have
again two parameters for the non-self-dual integrable version, and two
parameters $\la$ and $\phi=\vph$ for the self-dual case. For $\NC$ and $\beta=
\tilde{\beta} =1/\sqrt{2}$
the hamiltonian (\ref{H4}) coincides with that of the Ashkin-Teller
quantum chain \cite{kohmoto} for the special value $h= 1/3$ in the notations
of \cite{baake}. From the phase diagram of the Ashkin-Teller quantum chain we
know that in the case $h=1/3$ there is just one critical point at the self-dual
value $\la=1$. This is described by the ${\cal WA}_3$
rational model at $c=1$, which has an extended conformal symmetry with
fields of spin three and four.

\section{General results for the energy gaps of the $\ZED_n$-chiral Potts
model}
\par   Apart from the exact results for the non-chiral limiting case just
mentioned at the end
of the last section, there are many exact results for the superintegrable
case (\ref{H}), (\ref{aint}) (equivalent to (\ref{alpha}) with $\SD$): The
$\Zed_2$-case, which is the standard Ising quantum chain in a transverse field,
has been solved in refs.\ \cite{FrS,Kog,Pf}. Recently, as mentioned above,
also the complete spectrum of the $\Zed_3$-superintegrable quantum chain has
been calculated analytically \cite{DKCoy}. We shall first collect some
important partial results which are available for {\em generic} $\Zed_n$. \par
In order to state these, we shall define all gaps with respect to the lowest
$Q\!=\!0,\, p\!=\!0$-level, which we denote by $E_0(Q=0,p=0)$. Note that
$E_0(Q=0,p=0)$ is not necessarily
the ground state of the hamiltonian. Depending on the parameters chosen, the
ground state may be in any charge sector and may even
have non-zero momentum $p$, since the model contains incommensurate phases.
Nevertheless, in this section let us consider only gaps between
$p=0$-levels.
\par By $\Delta E_{Q,i}$ we denote the energy difference
\BEQ  \Delta E_{Q,i} = E_i(Q,p=0) - E_0(Q=0,p=0)        \label{gap} \EEQ
where $E_i(Q,p=0)$ is the {\em i}-th level $(i = 0,1,\ldots)$ of the charge
$Q$, momentum $p=0$-sector. In the $\Zed_2$-(Ising)-case of (\ref{H}) it is
well-known that the lowest gap $\Delta E_{1,0}$ has the
remarkable property of being linear in $\la$:
\BEQ  \Delta E_{1,0} = 2(1-\la).   \label{1} \EEQ
In 1983 Howes, Kadanoff and DenNijs \cite{how} found that the same is true
also for the self-dual version of the
$\Zed_3$-chiral Potts model, eq.(\ref{H3}) at the special chiral angle
$\SD$. They discovered this by calculating the high-temperature
expansion up to 10th order in $\la$, finding that for these special angles
all higher coefficients in the expansion, starting with the coefficient of
$\la^2$, are zero.
\par In \cite{geh} it was the desire to generalize (\ref{1}) so as to
obtain for {\em all}~ $\Zed_n$ and {\em all}~ $Q$ the simple formula
\BEQ \Delta E_{Q,0} = 2 Q(1-\la) \hspace{15mm} (\la<1,~~Q=1,\ldots,n-1)
                                                      \label{Qa} \EEQ
which lead to choose the values (\ref{aint}) for the coefficients $\alpha_k$
and $\bar{\alpha}_k$ and to find the superintegrability of the model.
In \cite{geh}
the linearity of the gaps was checked through numerical calculations.
The validity of (\ref{1}) for the superintegrable $\Zed_3$-model to all orders
of a perturbation expansion in $\la$ was first shown
in \cite{au2} using a recurrence formula. Later, using analytic methods,
Albertini {\em et al.} \cite{al2,au2} for the superintegrable $\Zed_3$-case,
and then Baxter \cite{ba1} for all superintegrable
$\Zed_n$, have calculated the lowest gaps of the
hamiltonian eq.(\ref{H}) in the thermodynamic limit $N \ra \infty$.
\par Using the definition (\ref{gap}), Baxter's analytic results for the
superintegrable model can be summarized in the form:
\BEA  \Delta E_{Q=1,0} & = & 2 (1-\la) \hspace{17mm} (\la < 1)
      \label{q1} \\   \Delta E_{Q=0,1}
    & = & 2 n (\la-1) \hspace{14mm} (\la > 1)~. \label{q2}    \EEA \par  We now
attempt to interpret the spectrum of (\ref{H}) and the gap formula (\ref{Qa})
in terms of particle excitations. Consider first eq.(\ref{Qa}). This would be
obtained if the model contained only one single particle species with mass
$m_1 = 2(1 -\la)$ and $\Zed_n$-charge $Q\!=\!1$. The lowest
$\Delta Q \neq 1$-gaps would then arise from the thresholds of the scattering
of $Q$ of these $Q\!=\!1$-particles.
\par    In order to check whether it is correct to describe the spectrum
at $\SD$ this way, or whether more fundamental particles must be present,
we shall study the zero momentum part of
the spectrum as a function of the chiral angles, when these
move away from the special superintegrable values. We shall start considering
the spectrum down at $\NC$, where we know the particle structure from the
thermally perturbed $\WA$-models and move then towards higher chiral angles.
\par The scaling regime around $\la=1$ of the $\WA$-models is known to
contain $n-1$ particle species, one in each of the non-zero
$\Zed_n$-charge sectors $Q=1,\ldots,n-1$. The ratios of their masses $m_Q$
are \cite{koeb}: \BEQ \frac{m_Q}{m_1}=\frac{\sin{(\pi Q/n)}}{\sin{(\pi/n)}},
\hspace{1cm} (Q=1,\ldots,n-1).  \label{koeb}  \EEQ      The particles
with masses $m_Q$ and $m_{n-Q}$ form particle-antiparticle-pairs.
Near $\la=1$ the mass scale $m_1$
behaves as  \BEQ  m_1 \sim (1-\la)^{(n+2)/2n}. \EEQ
as it follows from the conformal dimension $x=4/(n+2)$ of the leading
thermal operator of the $\WA$-model.
Looking at (\ref{koeb}), we can see that in the scaling $\WA$-model all
particles are {\em isolated} non-degenerate levels in the spectrum of
their respective charge sector. E.g. for the particle with mass $m_Q$ to be
on the scattering threshold of $Q$ of the lightest particles with mass $m_1$,
we must have $Q \sin{(\pi/n)}= \sin{(Q\pi/n)}$, which is possible only
in the limit $n \ra \infty$. Similarly, e.g. for $n\geq 5$, the $m_4$-particle
is below the threshold $m_2 + m_2$, and approaches this threshold only as
$n \ra \infty$.
\par We now want to follow these single particle levels in $\la$ to outside
the scaling region and as functions of the chiral angles
$\vph$ and $\phi$, in order to get information about what will happen at
$\vph,\phi \ra \pih$. Since near $\la=1$ the gaps can only be calculated
numerically for generic $\phi,\vph$ (we shall report such numerical
calculations
later), it is most simple to consider first the
small-$\la$-expansion for the lowest $p=0$-gaps of the hamiltonian (\ref{H}).
Using the coefficients in the form (\ref{alpha}), but not assuming
(\ref{int}), to first order in $\la$ we get:
\BEA  \Delta E_{Q,0}\!\!&\!=\!&\!\!  \left( \SK \bar{\alpha}_k(1-
 \omega^{Qk}) \right)~~~-\la(\alpha_Q + \alpha_{n-Q}) + \ldots   \nonumber \\
  &\!=\!&\!\! \left(\SK \frac{2 \sin{(\pi kQ/n)}}{\sin{(\pi k/n)}}
     \sin\left((\frac{2k}{n}\!-\!1)\vph+\frac{\pi kQ}{n}\right) \right)
-\frac{2\la}{\sin{(\pi Q/n)}}\cos\left( (\frac{2Q}{n}\!-\!1)\phi\right)
+\ldots   \nonumber    \\  & &  \label{ST}                  \EEA
Eq.(\ref{ST}) fulfils the CP-relation      \BEQ
\Delta E_{Q,i}(\la,\vph,\phi) = \Delta E_{n-Q,i}(\la,-\vph,-\phi)
\hspace*{12mm} (i=0,1,\ldots) \label{CPH} \EEQ     In the sum in
(\ref{ST}), the pairs of terms for $k$ and $n-k$ are equal.
\par For $\vph=0$ and for $\vph=\pih$
it is easy to simplify eq.(\ref{ST})
using the trigonometrical sum formulae \cite{RYG,HeLa}
\BEA  \SK \sin{(k\pi/n)} &\!=\!& \cot{\frac{\pi}{2n}}, \hspace*{1cm}
 \SK\sin^2{(k\pi Q/n)}~=~n/2, \nonumber \\ \SK\sin^3{(k\pi/n)}&\!=\!&
\frac{1}{4}\left(3\cot{\pn}-\cot{\frac{3\pi}{2n}}\right) \hspace*{8mm} etc.
\label{sumf} \EEA       For $\NC$ we obtain:      \BEQ \Delta E_{1,0} =
 2\cot{\pn}-\frac{2\la}{\sin{(\pi/n)}}+\ldots \label{W1} \EEQ
and \BEQ \Delta E_{2,0}=2\cot{\pn}+2\cot{\frac{3\pi}{2n}} - \frac{2\la}
 {\sin{(2\pi/n)}} +\ldots  \label{WAS}  \EEQ         We identify
\BEQ m_Q \equiv \Delta E_{Q,0}, \label{WW} \EEQ and conclude from (\ref{koeb})
and (\ref{WAS}) that the mass ratio $m_2/m_1$ must decrease when going from the
scaling region $\la \approx 1$ to $\la=0$. E.g. for $n=4$, eq.(\ref{koeb})
gives $m_2/m_1 = \sqrt{2}$, whereas at $\la=1$  from (\ref{W1}) and (\ref{WAS})
we get $m_2/m_1 \approx 1.17157$. Observe, that also according (\ref{WAS}),
$m_2$ stays below the $Q=2$-sector threshold at $2 m_1$, and generally, the
lowest levels of each charge sector remain isolated down to $\la=0$.
\par We now look how the perturbation formula (\ref{ST}) behaves in the
superintegrable limit. Inserting $\SD$,
we obtain  \BEQ \Delta E_{Q,0}=2\SK\sin^2{(\pi kQ/n)} -\SK\sin{(2\pi kQ/n)}
 \cot{(\pi k/n)} -2\la +\ldots  \label{zw}  \EEQ
The sums can be calculated explicitly, leading to the simple result
\BEQ \Delta E_{Q,0} = 2(Q -\la) + O(\la^2). \label{fasu}  \EEQ
Comparing (\ref{fasu}) to (\ref{Qa}), we have an apparent contradiction, since
in (\ref{fasu}) the factor $Q$ is {\em not} multiplying $\la$.    \par
Our detailed non-perturbative numerical analysis of the spectrum
for the $\Zed_3$- and $\Zed_4$-cases (to be discussed
in the next Section) shows that the single particle levels vary smoothly
when increasing $\vph$ and $\phi$ starting from $\NC$.
As soon as the chiral angles become non-zero, the
particle-antiparticle-pairs $m_Q$ and $m_{n-Q}$ split: $m_1$
decreases and $m_{n-1}$ increases as $\vph,\phi$ increase. Approaching
$\SD$, $m_2$ becomes twice as large as $m_1$, so that at $\SD$, $m_2$ sits just
at the $m_1+m_1$-scattering threshold. So, our non-degenerate perturbation
theory should break down for the channel $Q=2$ at
$\vph, \phi \ge \pih$ \cite{how}, and, indeed we see this in the calculation of
the $\la^2$-contribution to $\Delta E_{2,0}$. Because for general $\Zed_n$ the
explicit expression for the $\la^2$-term in (\ref{ST}) is too involved, here we
specialize and give the result for the self-dual $\Zed_3$-case. We find
\BEQ  \Delta E_{1,0}(\vph,\la) = 4 \sin\frac{\pi -\vph}{3} -
 \frac{4\la}{\sqrt{3}}\cos\frac{\vph}{3} + \la^2 f(\vph) +O(\la^3).
\label{Q1} \EEQ
where \BEQ f(\vph) =-\frac{1}{3\cos\vph} \left(
 2 \sin {\pi - 4 \vph \over 3}
- 4 \sin {\pi + 2 \vph \over 3} + 3\sqrt{3} \right).  \label{fph} \EEQ
$f(\vph)$ is smooth at $\vph=\pih$ (we have $f(\pih)=-\sqrt{3}/2$),
but it is singular for $\vph \ra -\pih$. Applying (\ref{CPH}), from (\ref{Q1})
we get $\Delta E_{2,0}$:
\BEQ  \Delta E_{2,0}(\vph,\la) = 4 \sin\frac{\pi +\vph}{3} -
\frac{4\la}{\sqrt{3}}\cos\frac{\vph}{3} + \la^2 f(-\vph)
+O(\la^3).\label{Q2}\EEQ
While formula (\ref{Q1}) for $\Delta E_{1,0}$ can be used in the whole range
$0 \le \vph < \pi$ (considering non-negative $\vph$), the corresponding
expression (\ref{Q2}) is valid only for $0 \le \vph < \pih$ and diverges at
$\vph=\phi=\pih$.
The $\la^2$-contribution to the analogous $\Zed_4$-formula
will be given in Sec.5. It shows a similar divergence in the expression
for $m_2$.
\par Since (\ref{Q2}) breaks down at $\vph = \pi/2$, we need further
information in order to decide whether the $Q=2$-particle survives
at and beyond $\vph = \pi/2$. Then this will clarify the question whether the
single particle picture described above makes sense at the superintegrable
line.

\section{Finite-size numerical calculation of the low-lying spectrum
of the $\ZED_3$-chiral quantum chain}
\subsection{Convergence exponent $y$}
\par We shall now describe several detailed numerical checks of the
two-particle picture from
the spectrum in the high-temperature massive region of the $\Zed_3$-model.
For this purpose, we have calculated numerically the eight lowest $p=0$-levels
of each charge sector of the hamiltonian
(\ref{H3}) for $N = 2,\ldots,12$ sites. While in the last Section, we
concentrated on the single particle levels, now we shall also look into the
scattering states and check, whether the corresponding thresholds appear
as expected.
\par   For chains of up to
$N=8$ sites we are able to diagonalize $H^{(3)}$ exactly. Using
Lanczos diagonalization we can use up to $N=12$ sites. We shall
always give the results for periodic boundary conditions, but we have
also partially checked the results with twisted boundary conditions.
The extrapolation to $N \rightarrow \infty$ is done using both the
Van-den-Broeck-Schwartz algorithm \cite{vbs} and rational approximants
\cite{bust} (for details on the application to quantum chains and
error estimation see \cite{HS}).
\par Neighbouring higher levels sometimes cross over
as functions of the number of sites $N$, so that for these there is a danger
of connecting wrong sequences. This can be controlled by calculating the
leading power of the convergence, $y$, defined by
\BEQ   \Delta E(N) - \Delta E(\infty) = N^{-y} + \ldots     \label{y1} \EEQ
and checking the smoothness of the approximants $y_N$:
\BEQ  y_N = - {\textstyle
 \ln (\frac{\Delta E(N)-\Delta E(\infty)}{\Delta E(N-1)-\Delta E(\infty)})
/ \ln (\frac{N}{N-1}) } \label{y2} \EEQ
In (\ref{y1}) and (\ref{y2}), $\Delta E(N)$ is a gap $\Delta E_{Q,i}$
calculated for $N$ sites.
\par   The calculation of $y_N$ is also very useful for distinguishing between
single particle levels and two- or more particle scattering states:
For single particle levels we expect
\BEQ   \Delta E(N) - \Delta E(\infty) = \exp(-N/\xi) +\ldots \label{y3} \EEQ
i.e.\ exponential convergence in $N$ ($\xi$ is a correlation length),
so that for these the $y_N$ should increase very fast with $N$.
In contrast to this, for two-particle states we should have a power behaviour
in $N$, more precisely, we should have
$\lim\limits_{N\rightarrow\infty} y_N = 2$.
\subsection{The spectrum for the superintegrable case $\vph = \pi/2$}
We first discuss our numerical results for
the $\Zed_3$-superintegrable case $\phi = \vph = \pi/2$.
Table~1 lists results for 16 low-lying gaps for $\la = 0.5$
in the high-temperature regime, where from (\ref{1}) we know that
$m_1 \equiv \Delta E_{Q=1,0} = 1$. As we have checked by repeating the
calculation for several other values of $\la<1$, this value $\la=0.5$
is not special regarding the structure of the spectrum. However, there is
the nice feature that because of (\ref{Qa}) for this $\la$ the gaps
should approach integer values in the limit $\NIF$.
\par  In the $Q=1$-sector we see an isolated lowest state at
$m_1 = 2(1-\la)$ followed by a bunch of levels at $4m_1$. Looking into Table~2
which gives the convergence with $N$ as parametrized by $y$ according
to (\ref{y1}), we see that indeed the $m_1$-level ($Q\!=\!1$; column $i=0$)
shows exponential convergence,
whereas the levels $Q\!=\!1$; $i\!=\!1$, $2$ converge as $N^{-3}$ and $N^{-2}$,
respectively. This indicates that the $i=1$-level is due to 3-particle
($m_1+m_1
+m_1$)-scattering, and the $i=2$-level due to
2-particle ($m_2+m_2$)-scattering as written in the bottom line in Table 1.
\par The $Q\!=\!2$-sector shows no isolated ground state, but starts with a
bunch of levels around $\Delta E =2$. There is strong evidence for
another bunch around $\Delta E =5$,
of which Table 1 shows just the first level. The convergence for the further
$\Delta E =5$-levels is poor and we do not give the numbers.
A clue to the nature of the $\Delta E =2$-levels is found through their
convergence for increasing $N$: Table 2 shows clearly that the level labelled
$Q\!=\!2; \ i\!=\!0$ has an $N$-dependence which drastically differs from that
of the other levels, indicating that this is the
level of a single particle with mass $m_2 =4(1-\la)$
(we have checked the $\la$-dependence repeating the calculation for six
other values of $\la$). So, we can answer the question posed at the end of the
last Section: The $m_2$-level can be clearly seen in the superintegrable case,
although it lies at the edge of the scattering threshold.
The power-behaved $Q=2$-states at $\Delta E=2$ should then be interpreted as
$m_1 +m_1$-states. The level $Q=2; \ i=5$ may be $m_1 +m_2 +m_2$, which has the
correct total $\Zed_3$-charge $Q=2$.
\par The pattern at $Q=0$, $\Delta E=3$ is as expected by this picture: There
should be $m_1 +m_2$-states and $m_1 +m_1 +m_1$-states, both distinguished
by different values of $y$, as observed in Tables 1 and 2.
\par For other values of $\la$ ranging from $\la = 0.2$ to $\la = 0.8$
we find precisely the same structure. Not surprisingly, above $\la =0.8$ the
convergence with respect to $N$ is getting poor, since the relevant mass scale
$m_1$ is vanishing as $\la\ra 1$ and so at $\la=0.8$ it is already quite small.
\subsection{The spectrum off the superintegrable line}
\par After having found a simple two-particle pattern in the superintegrable
case, we now check how this structure gets modified for $\vph, \phi \neq \pih$.
In order not to vary too many parameters,
we consider only the INT and SD cases defined at the end of Sec.2.
For the INT-case explicit formulae for the spectrum are not yet available
(for some first attempts, see \cite{roan}).
For the SD-case there may be no integrability at all. So, for the following,
presently there is no alternative to our numerical or perturbative methods.
\par We start with the SD case. In  Sec.3 we discussed already the perturbation
expansion of the lowest gaps to order $\la^2$. This indicated that,
increasing $\vph$ from
the Potts-value $\vph=0$ to $\vph = \pi/2$ for fixed $\la< 1$ the mass
$m_1$ decreases and $m_2$ increases until for $\vph=\pi/2$
we reach the value $m_2/m_1=2$. Above $\vph=\pi/2$
we then should have $m_2/m_1 > 2$ which makes it difficult to isolate the
$Q=2$-particle in the $m_1 +m_1$-continuum (in case it is still there at all).
Fig.2 shows the different patterns
which we expect in the three charge sectors for $\vph <\pi/2$ (left
hand side) and for $\vph > \pi/2$ (right). The extrapolation of the
finite-size numbers reported in
Table~3 confirms all details of these expected patterns. For
$\vph =2\pi/3$, among the three lowest $Q\!=\!2$ levels at $\Delta E = 2m_1$
we see no exponentially converging level. For $\vph > \pi/2$ all thresholds are
determined by $m_1$ alone, these are $3m_1$, $4m_1$ and $2m_1$ for $Q=0$, $1$,
$2$-sectors, respectively. \par In order to see, whether the $Q\!=\!2$-particle
survives in the $\vph > \pi/2$-region, we have to look for more and higher
levels. For {\em small} $\la$ we know from eq.(\ref{Q2}) (supposing this to be
still valid) where to search for $m_2$ at $\vph >\pi/2$, and so we have looked
slightly above the superintegrable line at $\vph = 7\pi/12$, $\la\leq 0.25$
among the 8 lowest levels for a fast converging one. Indeed, as it is shown in
Table~4, the level $i=6$ of the Table converges much faster than its
neighbours.
So, it is a good candidate for the $m_2$-level. However, with increasing
$\la$ this convergence diminishes quite fast
and an even/odd-$N$ hopping takes over. This may be due
to an increasing instability of the $m_2$-particle.
      \par  For $\vph <\pi/2$ the threshold in the $Q=0$-sector
depends also on $m_2$, being $m_1 +m_2$. The left column of Table 3 shows
that the
numerical values for the various thresholds come out very well and that e.g.
$m_2/m_1 = 1.47755$ at $\vph=\pi/4$ and $\la=0.5$. Unfortunately, the
determination of the corresponding values for $y$ is quite unsafe, since
for $\vph\neq\pih$ we have no exact values for $\lim\limits_{\NIF}\Delta E_i$
available as we had in the superintegrable case from (\ref{Qa}).
So the results we obtain for the $y_N$ depend strongly on the very unprecise
extrapolated results which are used for $\Delta E_i(\infty)$.
         \subsection{The integrable case for $\vph \neq \pi/2$}  \par
Turning now to the integrable case INT, our data in Table 5 show
the same pattern as in the SD case, only the convergence is less good
(for the sake of brevity we have omitted the finite-size data). It is
not clear whether the $\Delta E_{Q=2,0}$-level converges exponentially at all.
This result is somewhat surprising, because one might have expected a more
clear particle behaviour in the integrable case. We recall that nothing
is known about the integrability of the general SD case.
\par
There is not much difference between the SD-case Table 3 and the INT-case
Table 5. E.g. observe that in the right hand part of Table 5, which gives
an example for chiral angles above the superintegrable value,
the low-lying gaps come out clearly as integer multiples of the single scale
$m_1$, as it is expected for threshold values if $m_2 > 2m_1$.
The numerical precision is not very good because in the neighbourhood of the
incommensurate phase the finite-size approximants to e.g.\ $m_1$
first fall with $N$ and then rise again,
a property which is not easy to handle by usual extrapolation procedures.
\par    Apart from the just mentioned non-monotonous behaviour of the gaps
for increasing $N$,
the onset of the IC-phase is not felt in our study of the
$p=0$-levels even at $\vph = \phi = 5\pi/6$, $\la =0.5$. For still larger
angles, it becomes difficult and then practically impossible to arrange the
gaps for various $N$ into plausible sequences. Anyway, it is already
surprising that for $\vph = \pi/2$ and higher,
the low $p=0$-gaps do not seem to take any notice of the incommensurate
phase boundary, and vanish smoothly for $\la \ra 1$.
\par In the low-temperature regime $\la>1$ the pattern of the lowest levels
is the same in all three charge sectors: As expected, in the limit $\NIF$ the
ground state is three-fold degenerate (one isolated level in each charge
sector). Above the ground state we find for $\vph < \pi/2$ a gap which
is $\la( m_1(\la^{-1}) + m_2(\la^{-1}))$. Then we see
a quite dense sequence of levels starting
after a gap of $3\la m_1(\la^{-1})$ (the same gap in all three charge
sectors). If we would have normalized $H^{(3)}$ including a factor $\sqrt{\la}$
in the denominator of (\ref{H3}), the gaps would be just the same as those
in the $Q=0$ sector at the duality reflected value of
$\la$. This is a generalization of
Baxter's result (\ref{q2}) for the superintegrable case to $\vph<\pih$.

\section{Perturbative and finite-size numerical results for the
spectrum of the chiral $\ZED_4$-quantum chain}
\par In the last section we have reported our numerical evidence for a
two-particle picture of the spectrum of the off-critical chiral $\Zed_3$-Potts
quantum chain.
In this Section we shall show analogous numerical evidence
for a three-particle structure in the high-temperature regime of
the chiral $\Zed_4$-Potts quantum chain. \par We start giving
the high-temperature expansion of the lowest gaps of the self-dual
$\Zed_4$-quantum chain up to order $\la^2$. It reads explicitly:
\BEQ  \Delta E_{1,0} = 2 (1 + 2 \sin {\pi -2\vph \over 4} )
   - 2\la \sqrt{2} \cos(\vph/2) + \la^2 g(\vph) + \ldots   \label{z4a} \EEQ
\BEQ  \Delta E_{2,0} = 4 \sqrt{2} \cos(\vph/2)
                 - 2 \la + \la^2 h(\vph) + \ldots    \label{z4b}  \EEQ
\BEQ  \Delta E_{3,0} = 2 (1 + 2 \sin {\pi +2\vph \over 4} )
   - 2\la \sqrt{2} \cos(\vph/2) + \la^2 g(-\vph) + \ldots   \label{z4c} \EEQ
where $g$ and $h$ are the following functions:
\BEQ  g(\vph) =  \frac{1+\sqrt{2}\sin{(\vph/2)}}{2\sin{((2\vph+\pi)/4)}}
 +\frac{4\sin^2\vph +5\sin{\vph}-8\sqrt{2}\cos{(\vph/2)}-1}{4\sqrt{2}\cos{\vph}
  \cos{(\vph/2)}}      \EEQ
\BEQ h(\vph)=-\frac{2}{\cos{\vph}}
 \left(1-\sqrt{2}\cos{(\vph/2)} +2\cos^2{(\vph/2)} \right) \EEQ
The function $h(\vph)$ is singular for $\vph \ra \pih$, similarly, $g(\vph)$
is singular for $\vph \ra -\pih$. Of course, (\ref{z4c}) follows from
(\ref{z4a}) by a CP-transformation, see (\ref{CPH}). At $\NC$ the sectors
$Q=1$ and $Q=3$ are degenerate.   \par   At $\vph=5\pi/6$ and $\la=0$,
according to (\ref{z4a}) the gap $\Delta E_{1,0}$ vanishes, and above
this value of $\vph$ the ground state of the system is in the $Q=1$-sector.
This feature is due to the particular
choice $\beta=1/\sqrt{2}$ which we made after (\ref{H4}).
If we consider more general $\beta$, then this ground-state level
crossing moves to $\vph=\pi/2 + 2\sin^{-1}(\beta/\sqrt{2})$. So, for
$\beta\ge 1$ the ground state remains in the sector $Q=0$ up to $\vph=\pi$.
\par   For $\vph$ above its superintegrable value $\vph=\pih$ (with
$\beta=1/\sqrt{2}$) the scattering thresholds in all sectors are determined
by $m_1\equiv \Delta E_{1,0}$ alone, and
the "single-particle gaps" eqs.(\ref{z4b}) and (\ref{z4c}) move above the
scattering thresholds in the sectors $Q=2$ and $Q=3$, respectively. This
is quite analogous to what happened in the $Q=2$-sector of the $\Zed_3$-model.
\par
For all $\la>0$ the $Q=2$-gap is larger than both the $Q=1$- and $Q=3$-gaps.
For small values of $\la$ both $\Delta E_{1,0}$ and $\Delta E_{2,0}$
decrease with increasing $\vph$, while $\Delta E_{3,0}$ grows.
At $\vph=\pi/2$ the gaps are integer-spaced.
\par The high-temperature-expansion is reliable for small $\la$ and the lowest
levels only. So, for larger values of $\la$, we have calculated the six lowest
$p=0$-levels of each charge sector for the hamiltonian (\ref{H4}) numerically.
In the $\Zed_4$-case we are able to handle up to $N=10$ sites only.
Correspondingly, the extrapolations are less precise than in
the case of $\Zed_3$.                                     \par  First, we
present our results for some low-lying levels in the spectrum for the
superintegrable case of (\ref{H4}): $\SD,~ \beta=\tilde{\beta}=1/\sqrt{2}$.
Table 6 contains the results at $\la=0.5$, where from (\ref{Qa}) we should
have $m_1 = 1$, so that we expect the gaps for $\NIF$ to
approach simple integers. Indeed, this comes out well from our numbers.
As before, we have checked for four other values of $\la$ that
otherwise there is nothing special about choosing $\la = 0.5$.
\par In each of the $Q\neq 0$-sectors we see one level which shows very fast
convergence with $N$, and we identify this with the single-particle
state of mass $m_Q = Q$.       \par
In the charge sector $Q=0$ we can evaluate the four lowest gaps precisely
enough to assign them to $\Delta E = 4$. There should be four different types
of scattering states: $m_1\!+\!m_1\!+\!m_1\!+\!m_1$,~~$m_1\!+\!m_1\!+\!m_2$,~
{}~$m_2\!+\!m_2$ and $m_1\!+\!m_3$. While the exponents determine the first two
types, we would have to move away from $\SD$ in order to
distinguish the two latter states.
In the sector $Q=1$ we
see two excited states in addition to the $m_1$-particle state. They could
be one three-particle- and one two-particle scattering-state, both with
$\Delta E=5$. This is compatible with the expected states $m_1\!+\!m_2\!+\!m_2$
and $m_2\!+\!m_3$. \par The numerical convergence of the scattering levels
and the possibility of ordering the levels into
clear sequences in $N$ gets worse if we consider the sectors $Q=2$ and $Q=3$.
In the bottom line of Table 6 we give a few quite safe assignments.
\par
We now look numerically, how this particle pattern gets modified if we choose
values of the parameters $\phi\ne\pi/2$, $\vph\ne\pi/2$ for which no analytic
results are available.
In \mbox{Table 7} we have collected the first few lowest
energy gaps for three choices of the parameters.
\par   First, we look at values of $\phi = \vph < \pi/2$.
For $\phi = \vph = \pi/4$ (which is half-way to the $\cal W$-point)
and $\la = 0.5$ the masses of the $m_2$- and
$m_3$-states are clearly below the scattering thresholds and the remaining
levels can excellently be explained as multi-particle states. In the
charge-sector $Q=2$ there is one state with $\Delta E = 4.6$ which
seems to be unexplained by this pattern. However, the extrapolation
$\NIF$ is quite delicate and therefore the errors may be
too small. Thus, this state could also be a $2 m_1$-state.
\par  If we choose $\phi=\vph>\pi/2$, the low-lying levels of the
spectrum are given by $m_1$ alone. However, for small values of $\lambda$
and $\phi$ slightly above the super-integrable line the $Q=2$-charge-sector
shows a high level that converges very fast and is not an integral multiple
of $m_1$. The middle part of table 7 contains the explicit values for
$\phi = \vph = 3\pi/5$ and $\la = 0.10$. Here, the $m_2$-particle
is still clearly visible as a high level in the $Q=2$ sector.
Unfortunately, the $m_3$-particle is not visible in the $3 m_1$-continuum.
\par The right column of Table 7 shows that basically the same patterns
can also be observed in the INT case although here the approximation is less
good.     \par
To summarize, our numerical data supports a three-particle interpretation
of the low-lying spectrum for the chiral $\Zed_4$-Potts quantum chain
with $\vph$ in the range from $0$ to slightly above $\pi / 2$.
\section{Energy momentum relations for the single particle states}
\par  When interpreting the excitations of the model in terms of particles,
we should also look for their energy-momentum dispersion rule. On a lattice
with $N$ sites and periodic boundary conditions, the momentum $p$ can take
the $N$ values  \BEQ
p =-{\textstyle [\frac{N}{2}]},\ldots,-1,0,1,\ldots,{\textstyle [\frac{N}{2}]}.
        \label{p}     \EEQ
In calculating the limit $\NIF$ we use
\BEQ    P = \frac{2\pi}{N}\,p \hspace{15mm} (-\pi\leq P\leq\pi)  \EEQ  Except
for the $\WA$-case $\phi=\vph=0$ the hamiltonian (\ref{H}) does not conserve
parity, so, in general the curves $E(P)$ will not be symmetrical with respect
to $P \ra -P$.
\par  An expansion to first order in $\la$ gives a first orientation
of the energy-momentum relation near $\la=0$ for generic $\Zed_n$.
In an easy generalization of (\ref{ST}) we find for $Q \neq 0$:
\BEA \Delta E_{Q,0} & = &\left(\SK \bar{\alpha}_k (1-\omega^{Q k})\right)
   - \la (e^{i P} \alpha_Q + e^{-i P} \alpha_{n-Q}) +\ldots \nonumber  \\
   & = &   \left(\SK \bar{\alpha}_k (1-\omega^{Q k})\right)
   -\frac{2\la}{\sin{(\pi Q/n)}}\cos{(P -P_m)} +\ldots,
\label{pexp}    \EEA  where \BEQ P_m = (1 -2 Q/n)\phi. \label{Pmac} \EEQ
In the second line of (\ref{pexp}) we have inserted the definition
(\ref{alpha})
of $\alpha_k$ in the $\la$-dependent term. Since, to first order in $\la$,
the first term of (\ref{pexp}) contains no $P$-dependence, we see that
in the high-temperature limit, the violation of parity in the dispersion
relation of the particle with charge $Q$ is exclusively due to the presence
of a macroscopic momentum $P_m$ as given in (\ref{Pmac}). A particle and its
antiparticle feel macroscopic momenta which are opposite in sign.
\par In the parity conserving case $\NC$ the system can be made isotropic
between space and euclidian time rescaling the hamiltonian
by a suitable $\la$-dependent factor $\xi$.
So, for $\NC$ the particle will have the energy-momentum relation of the
lattice-Klein-Gordon equation
\BEQ    E/\xi = \sqrt{\mu_Q^2 + K^2},   \label{lKG} \EEQ
or \BEQ  E^2 = m_Q^2 + \xi^2 K^2,   \label{LKG}   \EEQ  where
$\xi$ is the rescaling factor, and we write
\BEQ   K=2\sin{(P/2)},      \hspace*{25mm}
m_Q=\mu_Q\xi.\EEQ \par The correct conformal normalization factor $\xi$ for
the $\WA$-quantum chains at the critical point $\la=1$ is
well-known\cite{Alc,GRR}:   \BEQ \xi(\la=1)=n. \label{CN} \EEQ     Using
formulae (\ref{pexp}) and (\ref{W1}),~(\ref{WAS}), the $m_Q$ and the
rescaling factors for the $Q=1$- and $Q=2$-particles of the $\WA$-chains near
$\la=0$ can be calculated explicitly. The $m_Q$ for $Q=1,2$ have been given in
(\ref{W1}), (\ref{WAS}) and (\ref{WW}), from which we find: \BEA
\Delta E_{1,0}\!&=&\!2\cot{(\pn)}-\frac{2\la}{\sin{(\pi/n)}}(1-K^2/2)
+\ldots \label{EP1} \\ \Delta E_{2,0}\!&=&\!2\cot{(\pn)}
  +2\cot{(\frac{3\pi}{2n})}
 -\frac{2\la}{\sin{(2\pi/n)}}(1-K^2/2)+\ldots   \label{EP12}   \EEA
from which we get different rescaling factors for particles
with different charge: Defining
\BEQ   \zeta =n\sqrt{\la} \label{ze} \EEQ
(coinciding for $\la=1$ with (\ref{CN})), for small $\la$ we obtain:  \BEQ
\xi_{Q=1}(\la) = \frac{\sqrt{2}}{n\sin{(\pn)}}~\zeta+\ldots \label{xi1} \EEQ
and \BEQ \xi_{Q=2}(\la) = \frac{\xi_{Q=1}(\la)}{\sqrt{2\cos^2\left(\pn
\right)-\frac{1}{2}}}+\ldots.         \label{xi2} \EEQ
Putting numbers, we find that the main variation of the $\xi_Q$ over the whole
range $0\le\la<1$ is determined (for low values of $n$) up to $10 \ldots 20 \%$
by the factor $\zeta$. E.g. for $\Zed_4$ at $\la\ra 0$ we have
\BEQ \lim_{\la\ra 0} \xi_{Q=1}/\zeta = 0.92388\ldots
\hspace*{15mm}\lim_{\la\ra 0}\xi_{Q=2}/\zeta = 0.84090\ldots
\EEQ similarly for higher $n$. For the Ising case $n=2$ we have exactly
$\xi_{Q=1}=\zeta$ for all $\la$ \cite{GI}.
\par  For $\NC$ the two parameters $\xi$ and $m_Q$
determine the dispersion curve exactly. E.g.\ in the ${\cal WA}_2$-case
($n=3$) we find for  $\la=0.00,~0.25,~0.50,~0.75,~0.90$:
\BEQ m_Q/(1-\la)^{5/6}
= 3.46410162,~3.58582824(1),~3.674214(2),~3.73(3),~3.8(1)
\label{pz1} \EEQ
and
\BEQ \xi/\zeta = 0.9428090,~0.9751702(1),~0.996697(2),~1.008(6),~1.00(1),
\label{pz2} \EEQ
respectively. At $\la=0.75$ and $\la=0.9$ we observe a non-monotonous
behaviour of the levels with $N$ which gives rise to a considerable
uncertainty in the extrapolation $\NIF$. At lower values of $\la$ the
exponential convergence clearly sets in already below $N=12$ sites, so that
there we may trust the extrapolation.
    \par    For $\la\ra 0$ the masses determining the
$N$-convergence go to infinity. So for $\la=0$ there is no $N$-dependence,
once a minimal value of $N$, dictated by the nearest neighbour-interaction, is
reached. This is in agreement with
(\ref{xi1}) and (\ref{xi2}) and shows that the high-temperature expansion
around $\la=0$ at the same time is also a nonrelativistic expansion of
(\ref{lKG}). For the ${\cal WA}_2$-case we have checked that the order
$\la^2$-term of (\ref{EP1}) agrees with the second order expansion term of
(\ref{lKG}): We find                               \BEQ   \Delta E_{1,0} =
\frac{2}{\sqrt{3}}\left(3-\la(2-K^2)-\la^2(1-K^2+K^4/6)+\ldots\right)  \EEQ
which fits to the form \BEQ \Delta E_{1,0}=m_1+(\xi K)^2/(2m_1)
-(\xi K)^4/(8m_1^3)+\ldots \EEQ   So, certainly to this order, the momentum
dependence of the ${\cal WA}_2$-single particle level $m_1$ is that of
the Klein-Gordon equation (\ref{lKG}) or (\ref{LKG}). \par
We have no physical interpretation why for low $n$ the ratio $\xi_Q/\zeta$
varies so little with $\la$ (in the $n=2$-case we have $\xi\equiv\zeta$),
          and we find it strange for the particle
interpretation that in general, $\xi_Q(\la)$ is $Q$-dependent.
\par  For the $\Zed_3$-superintegrable
case Albertini {\em et al.} \cite{al2} have given analytic
formulae and plots of the energy-momentum depencence of the $m_1$-particle
in the high-temperature
range $0 \leq \la < 1$, and for the first $Q\!=\!0$-excitation in the
low-temperature range $\la > 1$. Here we want to give simple approximate
expressions for the dispersion curves of both $Q=1$ and $Q=2$-particles,
not only for the superintegrable, but also for the
general self-dual $\Zed_3$-chiral model at $0\leq \la \leq 1$.
\par For $\vph, \phi>0$ we have no simple analytic form of the dispersion
equation and so we have performed fits to the curves $E(P)$ given by   \BEQ
E(P)= \sum_{m=0}^{3} a_{m}\cos^{m}(P-P_m) +\sum_{m=1,3}b_m\sin^m(P-P_m)
  \label{ex} \EEQ which come out good up to $\la \approx 0.6$.
Table 8 collects some of these fitted coefficients.    \par
For $\la < 0.5$ the convergence in $N$ is excellent for all $p$ if we use
up to $N=12$ sites.
For $\la \leq 0.1$ we can nicely fit
the data for the momentum dispersion of the $Q\!=\!1$-particle
just using eq.(\ref{pexp}).
For larger values of $\la$ inspection of the numerical curves
and the fits in Table~8 shows that
$P_m$ is decreasing with increasing $\la$, and in addition, the curves
start being unsymmetrical with respect to $P = P_m$. The larger the angle
$\vph$, and the larger $\la$ is the more terms in the expansion (\ref{ex})
are needed for a reasonable fit. Figs.3 - 5 show momentum distributions
for various values of $\la$ and $\vph$ together with our fitted curves.
As is seen from Fig.~5, for $\vph=2\pi/3$
and $\la=0.5$ more terms in the expansion would have been needed.
As we approach the incommensurate region, e.g.\ at $\vph=5\pi/6$, $\la=0.3$,
other levels get below the $m_1$ level at $\mid P \mid > \pi/2$, so that the
identification of the $m_1$-dispersion curve gets difficult. It may be that
in this region new particle types will show up.

\section{Conclusions}
We have shown that the low-lying spectrum of the massive high-temperature
regime of the self-dual $\Zed_n$-Potts quantum chain at low chiral
parameter values $\vph~ {}^{<}_{\sim}~ 2\pi/3$
can be described in terms of $n-1$ particles which carry $\Zed_n$-charges
$Q=1,\ldots,n-1$ (we denote their masses by $m_Q$).
This is seen by studying the variation of the single-particle masses from their
K\"{o}berle-Swieca-values at zero chirality $\vph=0$ up to and above the
superintegrable value $\vph=\pi/2$. For low inverse temperatures $\la$ we use
a perturbation expansion. For higher $\la$ we concentrate on the special cases
$n=3$ and $n=4$ and diagonalize the hamiltonian numerically. \par
In the $\Zed_3$-case the
mass ratio $m_2/m_1$ is shown to rise continously from $m_2/m_1=1$
at $\vph=0$ to $m_2/m_1=2$ for the superintegrable case $\vph = \pi/2$.
In the superintegrable case the $Q\!=\!2$-particle appears
precisely at the $m_1 +m_1$-scattering threshold. For $\vph \ra \pi$,
$m_1$ tends to zero. How far the $Q\!=\!2$-particle survives for $\vph >\pi/2$
is not clear. In Table~5 we give evidence that the $Q=2$-single particle level
is still present up to $\la=0.15$ and $\vph=7\pi/12$.
In the superintegrable case we identify two- and three-particle scattering
states through their different power behaviour in the chain
size $N$. The two single particle levels show exponential convergence
in $N$. The perturbation expansion of $m_2$ is shown to diverge at order
$\la^2$
for $\vph \ra \pi/2$.      \par   At small $\la$, the
nonconservation of parity in the model results only in the appearance of a
$Q$-dependent macroscopic momentum $P_m = (1-2Q/n)\phi$
in the energy-momentum dispersion relations of the particles, which is
$P_m=\pm\phi/3$ in the $\Zed_3$-case.
Our numerical data suggest that the average $P_m$ decreases with increasing
$\la$, perhaps $P_m \ra 0$ for $\la \ra 1$, but we have no simple
parametrization of the effects of parity violation at large $\la$ and restrict
ourselves to giving trigonometrical fit coefficients.     \par
Using the same methods we have also verified in detail that the
high-temperature
massive spectrum of the chiral $\Zed_4$-Potts quantum chain can be described
analogously in terms
of three massive particles with $\Zed_4$-charges $Q=1, \ 2$ and $3$.
In the superintegrable case the mass ratio equals the ratio of the charges,
such that now both the $Q=2$ and $Q=3$ particles appear at scattering
thresholds. As before, the scattering states can easily be identified by
their power behaviour in the chain length $N$, while the single particles
show exponential convergence with $N$.
\par
It would be interesting to use L\"{u}scher's \cite{LuW} method to obtain
numerical information on the phase-shifts or the $S$-matrix
from the $N$-dependence of the multi-particle states.
\vskip 2cm
\noindent
{\Large {\bf Figure captions}} \vskip 4mm    \par
Fig.~1:~~Schematic phase diagram of the self-dual $\Zed_3$-chiral Potts-model
as defined in eq.(\ref{H3}).       \\   \par
Fig.~2:~~Level structure for the self-dual $\Zed_3$-chiral Potts-model in the
high-temperature massive region.
Left hand side: for a chiral angle $\vph < \pi/2$ (below the superintegrable
value) and, right hand side: $\vph > \pi/2$ (above the
superintegrable value). In the latter case the thresholds are determined by
$m_1$ alone.                    \\    \par
Fig.~3:~~Energy-momentum relation of particle $m_1$ for the
$\Zed_3$-superintegrable case and various values of the inverse temperature
$\la$ in the high-temperature region. The small crosses and triangles
are finite-size values calculated for $N=6,\ldots,12$ sites. \\      \par
Fig.~4:~~Same as Fig.~3, but for $\vph$ below the superintegrable value
and both for $m_1$ (three dashed curves) and $m_2$ (full curve). \\  \par
Fig.~5:~~Same as Fig.~3, but for several values of $\vph$ above the
superintegrable value (towards the incommensurate region),
for the $Q=1$-particle $m_1$. The shift of the minima of the curves to
the right due to the macroscopic momentum $P_m$ is clearly seen.

\newpage   \noindent
{\LARGE {\bf Tables}}   \\ \vspace*{3mm} \par
\begin{tabular}{|r|ccccccc|} \hline
\multicolumn{1}{|c|}{ } & \multicolumn{7}{c|}{$\Delta E_{Q=0,i}$} \\
        $N$& $i=1$ & 2 & 3 & 4 & 5 & 6 & 7 \\  \hline
 7& 3.1598983& 4.8359585& 4.4520381& 5.9303676& 5.9032240&
7.0487030&6.3557797\\
 8& 3.1115649& 4.4240923& 4.2075367& 5.4868664& 5.5496714&
6.3987521&6.0338701\\
 9& 3.0808417& 4.1213280& 4.0168334& 5.1252880& 5.2384091&
5.8572524&5.7243697\\
10& 3.0604137& 3.8954662& 3.8661605& 4.8297326& 4.9700825&
5.4121174&5.4423701\\
11& 3.0463166& 3.7244809& 3.7455652& 4.5867795& 4.7406004&
5.0471104&5.1917195\\
12& 3.0362806& 3.5932058& 3.6478379& 4.3856960& 4.5446740&
4.7471645&4.9714669\\
\hline $\infty$&
 2.9998(3)& 2.986(7) & 3.01(1)  & 3.02(8)  & 3.05(6)  & 2.9(1)   & 2.95(8)   \\
\hline  $y$ &3.0(1)&2.9(2)&2.0(1)&2.1(1)&1.9(1)&2.8(6)&  ?     \\
   & $3m_1$ & $3m_1$ & $m_1+m_2$ & $m_1+m_2$ & $m_1+m_2$ & $3m_1$ & ? \\ \hline
\end{tabular}   \hzeile
\begin{tabular}{|c|ccccc|} \hline
\multicolumn{1}{|c|}{ } & \multicolumn{5}{c|}{$\Delta E_{Q=1,i}$} \\
        $N$& $i=0$ & 1 & 2 & 3 & 4 \\  \hline
 7& 0.9984874& 4.5173085& 5.7905491& 6.6049711& 6.1612566\\
 8& 0.9994289& 4.3760023& 5.4775432& 6.0785242& 5.8456347\\
 9& 0.9998044& 4.2809972& 5.2349161& 5.6757753& 5.5870148\\
10& 0.9999405& 4.2150750& 5.0442562& 5.3652765& 5.3749672\\
11& 0.9999851& 4.1680556& 4.8924532& 5.1235743& 5.2002368\\
12& 0.9999978& 4.1336962& 4.7700872& 4.9335093& 5.0552644\\
\hline $\infty$&
 1.0000000& 3.99(2)  & 4.00(1)  & 4.01(5)  & 4.01(3)     \\
\hline  $y$ &expon.&3.0(2)&2.0(1)&2.9(2)&1.9(1) \\
      & $m_1$ & $2m_1+m_2$ & $2m_2$ & $2m_1+m_2$ & $2m_2$ \\  \hline
\end{tabular}
\hzeile
\begin{tabular}{|r|cccccc|} \hline
\multicolumn{1}{|c|}{ } & \multicolumn{6}{c|}{$\Delta E_{Q=2,i}$} \\
        $N$& $i=0$ & 1 & 2 & 3 & 4 & 5 \\  \hline
 7& 1.9995726& 2.7901373& 4.3798245& 5.6633417&          & 6.1294869\\
 8& 1.9996761& 2.6389905& 4.0256344& 5.3106885& 6.0425834& 5.8468390\\
 9& 1.9998304& 2.5261089& 3.7351818& 4.9697753& 5.8323103& 5.6487379\\
10& 1.9999247& 2.4399607& 3.4970152& 4.6583108& 5.5766919& 5.5065400\\
11& 1.9999701& 2.3729391& 3.3008958& 4.3808694& 5.3110577& 5.4022604\\
12& 1.9999893& 2.3198948& 3.1384071& 4.1365977& 5.0527923& 5.3242970\\
\hline $\infty$& 2.0001(1)& 2.002(7) & 2.00(5) & 2.0(1) & 2.6(5) & 4.99(1) \\
\hline $y$ &expon.&2.0(1)&1.9(1)&1.9(2)&  ?   &2.9(2)\\
           &$m_2$& $2m_1$ & $2m_1$ & $2m_1$ & ? & $m_1+2m_2$ \\  \hline
\end{tabular}  \hzeile
Table~1: The lowest energy gaps $\Delta E_{Q,i}$, as defined in eq.(\ref{gap}),
for the $\Zed_3$-hamiltonian eq.(\ref{H3}) and the superintegrable case
$\SD$, $\la = 0.50$.
The numbers given in brackets indicate the estimated error in the last
written digit. All calculations of the gaps have been performed to 12 digit
accuracy and using all $N=2,\ldots,12$ sites for the extrapolation $\NIF$.
In order to save space, in the Tables we give less digits and omit the low-$N$-
data. Details on our determination of the exponent of convergence $y$, defined
in (\ref{y1}), (\ref{y2}) is given in Table 2 below. In the last line of each
Table above, we give our particle interpretation of the various levels, as
deduced from the result for $y$.
\newpage   \noindent
\begin{tabular}{|c|cccc|ccc|ccc|} \hline  \multicolumn{1}{|c|}{ } &
        \multicolumn{4}{c|}{$y_N$ for $\Delta E_{Q=0,i}$} &
        \multicolumn{3}{c|}{$y_N$ for $\Delta E_{Q=1,i}$} &
        \multicolumn{3}{c|}{$y_N$ for $\Delta E_{Q=2,i}$} \\
   $N$ & $i=1$ & 2 & 3 & 4 &$i=0$ & 1 & 2 &$i=0$& 1 & 2\\  \hline
  7& 2.6413 & 1.7475 & 1.2814 & 1.0998 & 5.6670  & 2.1812 & 1.3354 &        &
                                                            1.5124 & 1.0720 \\
  8& 2.6955 & 1.9024 & 1.3808 & 1.2290 & 7.2946  & 2.3893 & 1.4389 & 2.0775 &
                                                            1.5900 & 1.2068 \\
  9& 2.7348 & 2.0293 & 1.4594 & 1.3340 & 9.0948  & 2.4728 & 1.5230 & 5.4923 &
                                                            1.6503 & 1.3140 \\
 10& 2.7646 & 2.1348 & 1.5222 & 1.4212 & 11.2974 & 2.5376 & 1.5917 & 7.7025 &
                                                            1.6973 & 1.4013 \\
 11& 2.7879 & 2.2232 & 1.5731 & 1.4947 & 14.5079 & 2.5883 & 1.6482 & 9.6994 &
                                                            1.7340 & 1.4733 \\
 12& 2.8067 & 2.2976 & 1.6148 & 1.5573 & 21.8070 & 2.6287 & 1.6948 & 11.7909&
                                                            1.7633 & 1.5334 \\
\hline  $\infty$ &3.0(1)&
 2.9(2)&2.0(1)&2.1(1)&expon.&3.0(2)&2.0(1)&expon.&2.0(1)&1.9(1)\\  \hline
\end{tabular}
\hzeile
Table~2: Exponents $y_N$ of the convergence $N \ra \infty$ as defined in
eq.(\ref{y2}), for the $\Zed_3$-superintegrable case and $\la=0.5$ for the
lower
levels given in Table 1. "expon." means that the fast increasing
sequence of the $y_N$ indicates exponential convergence.
\zeile{1}
\begin{tabular}{|cc|l|l||cc|l|l|} \hline
\multicolumn{4}{|c||}{$\phi = \vph = \pi/4$, $\la = 0.5$} &
\multicolumn{4}{c|}{$\phi = \vph = 2 \pi/3$, $\la = 0.5$} \\ \hline
$Q$ & $i$ & $\Delta E_{Q,i}(\infty)$ & Particles &
      $Q$ & $i$ & $\Delta E_{Q,i}(\infty)$ & Particles \\ \hline
$0$ & $1$ & $3.9150(2)$   & $ m_1 + m_2$ &
      $0$ & $1$ & $1.75(1)$    & $3 m_1$    \\
$0$ & $2$ & $3.85(3) $    & $ m_1 + m_2$ &
      $0$ & $2$ & $1.80(5)$    & $3 m_1$    \\
$0$ & $3$ & $3.98(4) $    & $ m_1 + m_2$ &
      $0$ & $3$ & $1.8(1)$     & $3 m_1$    \\
$0$ & $4$ & $4.70(8) $    & $ 3 m_1$     &
      $1$ & $0$ & $0.5868(2)$  & $\equiv m_1$ \\
$0$ & $5$ & $3.90(5) $    & $ m_1 + m_2$ &
      $1$ & $1$ & $2.36(4)$    & $4 m_1$    \\
$1$ & $0$ & $1.5834366(1)$& $ \equiv m_1$  &
      $2$ & $0$ & $1.174(2)$   & $2 m_1$    \\
$1$ & $1$ & $4.68(1)$     & $ 2 m_2   $  &
      $2$ & $1$ & $1.18(2)$    & $2 m_1$    \\
$2$ & $0$ & $2.339607(1)$ & $ \equiv m_2$  &
      $2$ & $2$ & $1.19(5)$    & $2 m_1$    \\
$2$ & $1$ & $3.168(4)$    & $ 2 m_1   $  &
       \  &  \  &  \           &  \           \\
$2$ & $2$ & $3.17(5)$     & $ 2 m_1   $  &
       \  &  \  &  \           &  \           \\
\hline   \end{tabular}    \hzeile
Table~3: The lowest energy gaps $\Delta E_{Q,i}$ (extrapolated $\NIF$) of the
self-dual $\Zed_3$-model (\ref{H3}), for $\la=0.5$ and two values of $\vph$:
$\vph=\pi/4$ (below the superintegrable line), and $\vph=3\pi/2$ (above the
superintegrable line), together with their particle
interpretation.
\zeile{1}
\begin{tabular}{|c|cccc|cccc|} \hline  \multicolumn{1}{|c|}{ }
  &\multicolumn{4}{c|}{$\Delta E_{Q=2,i}\;(\vph=7\pi/12,\;\la=0.10$)} &
   \multicolumn{4}{c|}{$\Delta E_{Q=2,i}\;(\vph=7\pi/12,\;\la=0.15$)} \\
                                       \multicolumn{1}{|c|}{ }
  &\multicolumn{4}{c|}{$m_1=1.50335845$} &
   \multicolumn{4}{c|}{$m_1=1.41131611$} \\
 $N$ & $i=0$ & 5 & 6 & 7 &$i=0$ & 5 & 6 & 7 \\  \hline
 7&3.03919 &3.58623 &3.89205 &7.99416 &2.86777 &3.63650 &3.95076 &7.69924 \\
 8&3.03222 &3.72696 &3.89348 &7.90458 &2.85847 &3.85581 &3.98085 &7.57893\\
 9&3.02727 &3.65640 &3.89285 &7.83788 &2.85177 &3.75262 &3.96189 &7.49035\\
10&3.02363 &3.74332 &3.89313 &7.78699 &2.84678 &3.89547 &3.97910 &7.42310\\
11&3.02088 &3.69394 &3.89301 &7.74732 &2.84297 &3.81857 &3.96735 &7.37073\\
12&3.01875 &3.75171 &3.89306 &7.71579 &2.83999 &3.91801 &3.97795 &7.32905\\
\hline   \end{tabular}   \hzeile
Table~4: Selected $Q=2$-gaps for $\vph = 7\pi/12$ (slightly above the
superintegrable line) of the self-dual $\Zed_3$-model.
The numbering of the levels ($i=0,\ldots$) refers to $N=12$ sites, for a
smaller
number of sites there are less levels between $i=0$ and $i=5$. The values of
$m_1$ are very well determined because of fast convergence.
Observe that the $i=0$-levels converge excellently to $2m_1$. In both cases
level $i=6$ converges much faster with $N$ than the other levels, which may
indicate that this is the $m_2$-level. Non-monotonous behaviour
in $N$ is common in the neighbourhood of the IC-phase.
\newpage
\begin{tabular}{|cc|l|l||cc|l|c|} \hline
\multicolumn{4}{|c||}{$\vph = \pi/ 4$, $\phi = 19.47\ldots^{\circ}$} &
\multicolumn{4}{c|}{$\vph = 2\pi/3$, $\phi=135.58\ldots^{\circ}$} \\
\multicolumn{4}{|c||}{$\la = 0.75$} &
 \multicolumn{4}{c|}{$\la = 0.70$} \\     \hline
$Q$ & $i$ & $\Delta E_{Q,i}(\infty)$ & Particles &
      $Q$ & $i$ & $\Delta E_{Q,i}(\infty)$ & Particles \\  \hline
$0$ & $1$ & $1.933(2)$ & $m_1 + m_2$ &
      $0$ & $1$ & $1.50(1)$    & $3 m _1$ \\
$0$ & $2$ & $1.9(2)$   & $m_1 + m_2$ &
      $0$ & $2$ & $1.53(5)$    & $3 m _1$ \\
$0$ & $3$ & $1.8(3)$   & $m_1 + m_2$ &
      $1$ & $0$ & $0.495(2)$   & $\equiv m_1$ \\
$1$ & $0$ & $0.750(1)$ & $\equiv m_1$ &
      $1$ & $1$ & $1.99(1)$    & $4 m _1$ \\
$1$ & $1$ & $2.41(2)$  &$2 m_2$ &
      $2$ & $0$ & $0.995(1)$   & $2 m _1$ \\
$2$ & $0$ & $1.19(2)$  & $\equiv m_2$ &
      $2$ & $1$ & $1.00(1)$    & $2 m _1$ \\
$2$ & $1$ & $1.50(3)$  & $2 m_1$ &
      $2$ & $2$ & $1.04(5)$    & $2 m _1$ \\
 \  & \   &     \      &     \     &
      $2$ & $3$ & $2.52(5)$    & $5 m _1$ \\
\hline  \end{tabular}   \hzeile
Table~5: Lowest energy gaps $\Delta E_{Q,i}$ as in Table 3, but for two
different choices of the parameters in the {\em integrable} $\Zed_3$-model
case,
in which $\vph$ and $\phi$ are related by (\ref{int}).
On the left side we give an example with chiral angles below the
superintegrable
case, on the right hand side $\phi$,$\vph$ are taken above their
superintegrable
values. The pattern observed here is very similar to that of the self-dual case
in Table~3.
\zeile{1}     \noindent
\begin{tabular}{|r|cccc|ccc|} \hline
\multicolumn{1}{|c|}{ } & \multicolumn{4}{c|}{$\Delta E_{Q=0,i}$}
                        & \multicolumn{3}{c|}{$\Delta E_{Q=1,i}$} \\
 $N$ & $i=1$        & $2$          & $3$          & $4$          &
       $i=0$        & $1$          & $2$    \\  \hline
$ 4$ & $ 4.4870349$ & $ 6.6401198$ & $ 6.9386000$ & $ 7.1202527$ &
       $ 0.9782260$ & $ 6.2907501$ & $ 8.8182343$  \\
$ 5$ & $ 4.2443835$ & $ 6.2079844$ & $ 6.3448347$ & $ 6.4390786$ &
       $ 0.9932173$ & $ 5.7339386$ & $ 7.9355672$  \\
$ 6$ & $ 4.1352996$ & $ 5.6907770$ & $ 5.9280172$ & $ 6.0208759$ &
       $ 0.9985794$ & $ 5.4429212$ & $ 7.2245463$  \\
$ 7$ & $ 4.0807817$ & $ 5.3191415$ & $ 5.5469396$ & $ 5.7278875$ &
       $ 0.9999564$ & $ 5.2809826$ & $ 6.8098311$  \\
$ 8$ & $ 4.0511625$ & $ 5.0477033$ & $ 5.2606178$ & $ 5.2880905$ &
       $ 1.0001440$ & $ 5.1860132$ & $ 6.4550409$  \\
$ 9$ & $ 4.0339657$ & $ 4.8460464$ & $ 4.9582447$ & $ 5.0427288$ &
       $ 1.0000927$ & $ 5.1277432$ & $ 6.1875836$  \\
$10$ & $ 4.0234306$ & $ 4.6936534$ & $ 4.7248803$ & $ 4.8744672$ &
       $ 1.0000390$ & $ 5.0905398$ & $ 5.9834045$  \\
\hline
$\infty$&$4.0000(2)$&  $3.89(2)$   &  $ 3.9(9)$   &  $3.7(4)$    &
        $1.0001(1)$ & $4.97(3)$    & $ 5.7(6)$   \\
\hline
$y$  & $4.0(2)$     & $2.1(3)$     & $ 2.9(4)$    & $1.6(1)$     &
       expon.\      & $3.5(3)$     & $ 1.8(1)$   \\
  & $4m_1$ & $2m_2$ or & $2m_1+m_2$ & $2m_2$ or & $m_1$ &  ?  & $m_2+m_3$ \\
     & & $m_1+m_3$ & & $m_1+m_3$ & & & \\
\hline  \end{tabular}    \hzeile
\begin{tabular}{|r|cc|ccc|} \hline
\multicolumn{1}{|c|}{ } & \multicolumn{2}{c|}{$\Delta E_{Q=2,i}$}
                        & \multicolumn{3}{c|}{$\Delta E_{Q=3,i}$} \\
 $N$ & $i=0$       & $1$         & $i=0$       & $1$         & $2$ \\ \hline
$ 4$ & $1.9760981$ & $3.6344881$ & $3.0090309$ & $5.0706873$ & $5.4135243$ \\
$ 5$ & $1.9885583$ & $3.1951141$ & $2.9978592$ & $4.4899244$ & $4.9297443$ \\
$ 6$ & $1.9959972$ & $2.9027708$ & $2.9980224$ & $4.1048393$ & $4.5520862$ \\
$ 7$ & $1.9989351$ & $2.7013088$ & $2.9991277$ & $3.8412387$ & $4.2636142$ \\
$ 8$ & $1.9998278$ & $2.5582804$ & $2.9997209$ & $3.6558602$ & $4.0432463$ \\
$ 9$ & $2.0000213$ & $2.4539476$ & $2.9999379$ & $3.5222475$ & $3.8733462$ \\
$10$ & $2.0000330$ & $2.3759111$ & $2.9999964$ & $3.4237046$ & $3.7406336$ \\
\hline
$\infty$&$2.00001(3)$& $1.9(1)$  & $3.00001(6)$&  $2.8(2)$   & $ 3.02(4)$  \\
\hline
 $y$ & expon.\     & $1.9(1)$    & expon.\     & $2.1(2)$    & $1.7(2)$    \\
     &  $m_2$ & $2m_1$ & $m_3$ & $m_1+m_2$ & $m_1+m_2$ \\
\hline   \end{tabular}   \hzeile
Table~6: The lowest energy gaps $\Delta E_{Q,i}$, as in Table~1, but here
instead of the $\Zed_3$-case now
for the $\Zed_4$-hamiltonian eq.(\ref{H4}). Superintegrable case
$\phi = \vph = \pi/2$ for $\la = 0.5$. Note the overshooting in the
approximants for $m_1$ and $m_2$.
\newpage    \noindent
\begin{tabular}{|cc|l|l||cc|l|l||cc|l|l|} \hline
\multicolumn{4}{|c||}{$\phi = \vph = \pi/4$} &
 \multicolumn{4}{c||}{$\phi = \vph = 3 \pi/5$} &
  \multicolumn{4}{c|}{$\vph = \pi/4$, $\phi = 19.47\ldots^{\circ}$} \\
\multicolumn{4}{|c||}{$\la = 0.50$} &
 \multicolumn{4}{c||}{$\la = 0.10$} &
  \multicolumn{4}{c|}{$\la =0.75$} \\       \hline
$Q$ & $i$ & $\Delta E_{Q,i}(\infty)$ & Particles &
      $Q$ & $i$ & $\Delta E_{Q,i}(\infty)$ & Part. &
          $Q$ & $i$ & $\Delta E_{Q,i}(\infty)$ & Particles \\ \hline
$0$ & $1$ & $5.77(2)$     & $m_1 + m_3$  &
      $0$ & $1$ & $5.1(2)$     & $4 m_1$    &
          $0$ & $1$ & $2.767(4)$   & $m_1 + m_3$ \\
$0$ & $2$ & $5.76(8)$     & $m_1 + m_3$  &
      $1$ & $0$ & $1.2116781$  & $\equiv m_1$  &
          $0$ & $2$ & $3.2(1)$   & $2 m_1 + m_2$ \\
$0$ & $3$ & $7.1(2)$      & $2 m_2$      &
      $1$ & $1$ & $6.1(3)$     & $5 m_1$    &
          $1$ & $0$ & $0.8663(1)$ & $\equiv m_1$   \\
$1$ & $0$ & $2.03383(1)$ & $\equiv m_1$   &
      $2$ & $0$ & $2.42334(7)$ & $2 m_1$     &
          $1$ & $1$ & $3.50(1)$ & $2 m_1 + m_3$   \\
$1$ & $1$ & $7.1958(4)$   & $m_2 + m_3$  &
      $2$ & $1$ & $2.430(6)$   & $2 m_1$     &
          $2$ & $0$ & $1.529(2)$ & $\equiv m_2$   \\
$1$ & $2$ & $6.98(7)$     & $m_2 + m_3$  &
      $2$ & $5$ & $3.211450(1)$&$\equiv m_2$ &
          $2$ & $1$ & $1.8(1)$ & $2 m_1$   \\
$2$ & $0$ & $3.4282(1)$  & $\equiv m_2$   &
      $3$ & $0$ & $3.6349(5)$  & $3 m_1$     &
          $3$ & $0$ & $1.786(2)$ & $\equiv m_3$   \\
$2$ & $1$ & $4.15(6)$     & $2 m_1$      &
          & \   &     \        &     \         &
          $3$ & $1$ & $2.52(7)$ & $3 m_1$   \\
$2$ & $2$ & $4.59(1)$     &     ?          &
          & \   &     \        &     \         &
              & \   &     \        &     \         \\
$2$ & $3$ & $7.62(7)$     & $2 m_3$      &
          & \   &     \        &     \         &
              & \   &     \        &     \         \\
$3$ & $0$ & $3.7501(1)$   & $\equiv m_3$       &
          & \   &     \        &     \          &
              & \   &     \        &     \         \\
$3$ & $1$ & $5.458(2)$    & $m_1 + m_2$  &
          & \   &     \        &     \          &
              & \   &     \        &     \         \\
\hline  \end{tabular}   \hzeile
Table~7: The left and middle parts of this Table show a selection of lowest
energy gaps of the self-dual, but not super-integrable
$\Zed_4$-hamiltonian for two different choices of the parameters $\phi = \vph$,
with $\beta=\tilde{\beta}=1/\sqrt{2}$ and different $\la$. The right part of
the Table contains an example of the INT case. For most levels in this Table
the determination of the convergence exponent $y$ is very unsafe due to the
considerable uncertainty in $\Delta E_{Q,i}(\infty)$. So, as also in Tables 3
and 5, the
particle content is inferred almost exclusively from $\Delta E_{Q,i}(\infty)$
alone.     \zeile{2}
\begin{tabular}{|cc|cccccc|c|} \hline
$\vph$&$\la$& $a_0$& $a_1$& $a_2$& $a_3$&  $b_1$& $b_3$&$\vph_m$\\
$[deg]$ & & & & & & &  & $[deg]$ \\     \hline
 45.&0.25& 2.8745 &-0.5924 &-0.0549 &-0.0075  & 0.0033 &        & 39.\\
 45.&0.50& 3.0302 &-1.1653 &-0.2355 &-0.0764  &-0.0134 &        & 30.\\
 45.&0.75& 3.3502 &-1.6702 &-0.6028 &-0.2300  & 0.0864 &-0.2121 & 18.\\
 90.&0.25& 2.0436 &-0.5118 &-0.0412 &-0.0848  &-0.0112 &        & 81.\\
 90.&0.50& 2.2363 &-1.1331 &-0.2381 & 0.0014  &-0.0188 &        & 69.\\
 90.&0.70& 2.4363 &-1.4587 &-0.4244 &-0.1190  &-0.0757 &-0.0112 & 57.\\
 90.&0.85& 2.6161 &-1.5982 &-0.5752 &-0.2791  &-0.1981 &-0.0745 & 39.\\
120.&0.25& 1.4317 &-0.5570 &-0.0552 &-0.0022  &-0.0263 &        &105.\\
120.&0.50& 1.5407 &-0.8112 &-0.0909 &-0.2517  &-0.3291 &        & 57.\\
150.&0.20& 0.7281 &-0.4696 &-0.0028 & 0.0451  & 0.0013 &        &150.\\
171.&0.10& 0.2257 &-0.2260 &-0.0128 &-0.0009  &        &        &171.\\
\hline 
 45.&0.50& 3.7927 &-1.1973 &-0.1501 &-0.0874  &-0.3223 &        &-45.\\
\hline  \end{tabular}
\hzeile
Table~8: Coefficients of fits to the energy-momentum relation eq.(\ref{ex}) for
various $\vph$ and $\lambda$, all for the self-dual model. Except for the last
line, which is for the lowest $Q\!=\!2$-gap, all other fits are for the lowest
$Q\!=\!1$-gaps. The quality of the fits can be judged from Figs.~3,~4 and 5.
In the last column we quote $\vph_m$, which we define by $P_m = \vph_m/3$.

\end{document}